# Data journeys in popular science:
## Producing climate change and COVID-19 data visualizations at *Scientific American*

Kathleen Gregory*[1], Laura Koesten[2], Regina Schuster[2], Torsten Möller[2,3], Sarah Davies[4]

[1]Centre for Science and Technology Studies, Leiden University
[2]Faculty of Computer Science, University of Vienna
[3]Research Network Data Science, University of Vienna
[4]Department of Science and Technology Studies, University of Vienna

*Corresponding author

## Author Roles:

Conceptualization: KG, LK; Data Curation: KG; Formal Analysis: KG, LK, RS; Funding Acquisition: KG, LK, TM; Investigation: KG, LK; Methodology: KG, LK; Project Administration: KG; Resources: TM; Supervision: TM, SD; Visualization: KG, RS; Writing – Original Draft: KG; Writing – Review & Editing: KG, LK, RS, TM, SD

## Abstract

Vast amounts of (open) data are increasingly used to make arguments about crisis topics such as climate change and global pandemics. Data visualizations are central to bringing these viewpoints to broader publics. However, visualizations often conceal the many contexts involved in their production, ranging from decisions made in research labs about collecting and sharing data to choices made in editorial rooms about which data stories to tell. In this paper, we examine how data visualizations about climate change and COVID-19 are produced in popular science magazines, using *Scientific American,* an established English-language popular science magazine, as a case study. To do this, we apply the analytical concept of '*data journeys*' (Leonelli, 2020) in a mixed methods study that centers on interviews with *Scientific American* staff and is supplemented by a visualization analysis of selected charts. In particular, we discuss the affordances of working with open data, the role of collaborative data practices, and how the magazine works to counter misinformation and increase transparency. This work provides an empirical contribution by providing insight into the data (visualization) practices of science communicators and demonstrating how the concept of *data journeys* can be used as an analytical framework.

**Keywords:** data visualization, science communication, data practices, open data, data reuse, climate change, COVID-19

## Media Summary

Vast amounts of data are increasingly used to make arguments about crisis topics such as climate change and global pandemics. Data visualizations are central to bringing these viewpoints to broader publics, where 'narrative visualizations' are used in popular science to inform, entertain, and persuade. At the same time, data visualizations conceal the many contexts involved in their production, ranging from decisions made in research labs about how to collect data to choices made in editorial rooms about which stories to tell. Revealing the contexts involved in producing data and visualizations is important to facilitate data understanding and reuse, counter misinformation, and increase transparency.

In this paper, we apply the analytical concept of '*data journeys*' (Leonelli, 2020) to examine how data visualizations about climate change and COVID-19 are produced in popular science magazines, using *Scientific American,* an established English-language popular science magazine, as a case study. We use the idea of *data journeys* to present the results of a mixed methods study centering on interviews with *Scientific American* staff, which is supplemented by a visualization analysis of selected charts. We also incorporate 'visual vignettes' into our analysis, which include example charts provided with permission from *Scientific American.*

Our findings highlight one set of practices for making *data journeys* visible and more transparent in popular science while also critically exploring how data about climate change and COVID-19 are reused, visualized, and transported outside of academia to broader readerships. In particular, we highlight affordances of working with open data, the role of collaborative data practices, and how the magazine works to counter misinformation and increase transparency. Our work provides an empirical contribution by providing insight into the data (visualization) practices of science communicators and demonstrating how the concept of *data journeys* can be used as an analytical framework.

## 1. Introduction

Open data are increasingly used to communicate urgent messages about crisis topics, such as climate change and global pandemics, by both advocates and skeptics (Lee et al., 2021). Data visualizations are central to bringing these viewpoints to broader audiences (Harkness, 2020; Zhang et al., 2021). However, visualizations often conceal the many contexts involved in their production (Lynch, 1988; Pandey et al., 2014), ranging from decisions made in research labs about how to collect and share data to choices made in editorial rooms about which data stories to tell (Beaulieu & Leonelli, 2021; Cabreros, 2021).

Revealing the contexts involved in producing data and visualizations is important to facilitate data reuse (Gregory et al., 2020), increase transparency (Kennedy et al., 2020), and counter misinformation (King, 2022). Such contexts can be hinted at using a reference to a data source or dataset, although this is far from being a standardized practice in scholarly or science communication (Gregory et al., 2023; Zamith, 2019). A simple reference also does not represent the complexity of a *data journey* (Leonelli & Tempini, 2020). In such journeys, data travel from their sites of creation to places in which they are read and interpreted (Beaulieu & Leonelli, 2021). Each site through which data travel has its own norms and ways of working with data (Borgman, 2015; Edwards et al., 2011) and for producing scientific knowledge for different audiences (Burri, 2012; Burri & Dumit, 2008).

In this paper, we draw on the analytical concept of *data journeys* (Leonelli, 2020) to examine how data visualizations about climate change and COVID-19 are produced in popular science magazines, taking *Scientific American,* an established English-language popular science magazine, as a case study. Specifically, we explore the following research questions: What are the practices of producing data visualizations at *Scientific American*? How do science communicators work with data, and what do 'open data' enable? To address these questions, we present the results of a mixed methods study consisting of semi-structured interviews with the *Scientific American* staff and an analysis of a sample of visualizations and documentation provided by the magazine.

We begin by situating our study within the context of literature on the use of open data and visualizations in science communication, before detailing our conceptual framework. We then use this framework to structure our findings. We conclude by discussing how (open) data and visualization production practices allow, shape, and inhibit data journeys in the popularization of science along four dimensions which we identify in our data: i) the affordances of open data; ii) the role of collaborative work; iii) misinformation; and iv) transparency.

## 2. Background and conceptual framework

### 2.1. Open data in science communication

Increasing amounts of data are being shared online, openly when possible (Benjelloun et al., 2020; Tenopir et al., 2020). Data are shared for many reasons, ranging from increasing transparency and reproducibility to enabling others to reuse and combine data in new ways (Pasquetto et al., 2019; Zuiderwijk et al., 2015). Although one of the primary aims of the broader open science movement is to make scientific information and data available to non-specialists, the practices of scientists are typically foregrounded, rather than those of decision-makers or journalists (Elliott, 2022). Science communicators do reuse open data, e.g., in stories about crisis topics such as climate change and the COVID-19 pandemic (Newell et al., 2016; Zhang et al., 2021); work examining *how* they do so, however, particularly to create data visualizations, is relatively sparse.

In a recent review of the use of open data in journalism before and during the COVID-19 pandemic, Fleerackers and colleagues (2023) highlight that the majority of data journalists make use of data from governmental sources rather than 'open research data,' which they define as publicly available data created at universities or research institutions. Stalph and Heravi (2021) find a similar pattern for data visualizations in award-winning journalistic pieces, where only 4% of visualizations relied on data from academic institutions, perhaps reflecting the claim that research data are more relevant and interpretable to scholars than to journalists or other audiences (Elliott & Resnik, 2019; Fleerackers et al., 2023).

Recent work analyzes the (open) data sources used to create visualizations about both COVID-19 and climate change specifically. Zhang and colleagues identified 450 different data sources used to create visualizations about COVID-19, although they did not classify the types of data sources used (Zhang et al., 2021). Governmental and intergovernmental data sources play a particularly important role in climate science (Edwards, 1999; Stanhill, 2001) and in visual data communication to public audiences. This is evidenced in part by the proliferation of visualizations in legacy and social media that have been derived from reports by the Intergovernmental Panel on Climate Change (IPCC) (O'Neill et al., 2015; Schuster et al., 2023).

## 2.2. Data visualizations in science communication and popular science magazines

Data visualizations, also termed charts, are commonly used within science communication to present data in ways that are designed to be understandable and engaging to different audiences (Böttinger et al., 2020). Although a range of visualization types are mobilized for these purposes (Borkin et al., 2016; Franconeri et al., 2021), visualizations within popular science tend to be 'narrative visualizations,' which are designed to convey stories, or accounts of events, facts, and the connections between them (Segel & Heer, 2010). Simplification (of data, visualization designs, and textual elements) is seen to be a key part of visual storytelling for nonacademic audiences (Engebretsen, 2020; Schuster et al., 2023).

Narrative visualizations are often embedded within other text or exist in online environments, where readers' attention is directed through the use of visual characteristics, chart placement, and interactive elements (Segel & Heer, 2010). To facilitate navigating and interacting with visualizations, producers use rhetorical techniques at different editorial layers (Hullman & Diakopoulos, 2011). Data visualizations in science communication are typically studied by analyzing articles in general news sources, such as *The New York Times, The Guardian,* or *Zeit Online.* Such studies focus on characterizing the number, complexity, and types of visualizations within pieces of data journalism (Knight, 2015; Stalph, 2018; Stalph & Heravi, 2021; Tandoc & Oh, 2017). Data visualizations are also a key element in popular science magazines, although scholarship in this area has typically focused on other visual forms, specifically photography (Born, 2020; Remillard, 2011).

Some visualizations in popular science magazines are reworked and 'recontextualized' versions of charts originally published in academic journals (Heekeren, 2021), although the division between what constitutes an academic journal and a popular science magazine is not always clear cut. The multidisciplinary peer-reviewed academic journals *Nature* and *Science* also publish journalistic pieces and engage in different levels of visual design intervention, depending on an article's intended audience and copyright concerns. At the journal *Nature,* designers make only minor alterations to visualizations in peer-reviewed scientific articles while they create more bespoke graphics to use in news stories (Krause, 2017).

Popular science magazines are perhaps unique in their use of data, although this has yet to be explicitly investigated. We suggest that these magazines likely engage with data that have passed through research settings rather than governmental data sources alone. Such data may have been encountered through the academic literature as charts are recontextualized (as in Heekeren, 2021) or through contact with scientists, who contribute to pieces (Bowler, 2013). Popular science magazines also occupy a unique corner in the science communication landscape, as hybrid spaces between academic journals and the mass media, where "scientific truth claims are selected, recontextualized and actively transformed" (Born, 2020, p. 34), making them an interesting case for analysis using the *data journeys* concept.

## 2.3. Data visualization production practices in science communication

Much critical work studying practices of how data visualizations are produced stems from a chart-based approach, where analysis begins with a close reading of visualizations to identify and question traces of production decisions (Klein, 2022; Poirier, 2021; Sorapure, 2019). The chart-based approach provides valuable insights, especially when visualization producers are distant in time or space. Visualizations can also be studied at their sites of production, which enables exploring the practices of how visualizations are actually constructed (Burri & Dumit, 2008).

Ethnographically-oriented approaches provide insight into the broader data practices of journalists (e.g., Borges-Rey, 2017; Parasie, 2015; Parasie & Dagiral, 2013). In their observations of investigative journalists, Parasie (2015) identifies two paths for how reporters work with data: a *hypothesis-driven path*, where journalists seek data to support or invalidate existing leads, and a *data-driven path*, in which journalists follow a 'light hypothesis' to explore data and develop new directions. Drawing on sixty interviews of journalists across Europe, Kennedy and colleagues explicitly explore data visualization practices in newsrooms (Kennedy et al., 2020; Weber et al., 2018). Although journalists in these studies describe linking to and crediting data sources as a way to increase transparency, such processes are time-consuming and not implemented in a standard way across newsrooms (Kennedy et al., 2020).

### 2.4. Conceptual framework: Data journeys and the popularization of science

The processes of working with (open) scientific data and designing visualizations in popular science are often investigated separately. We argue that these practices are not isolated from each other, but that data visualizations in popular science result from intersecting contexts and practices within and between science and journalism. Combining the concepts of data journeys and scientific popularization into a conceptual framework (Figure 1) allows us to explore these intersections and examine the role of popular science magazines in data journeys.

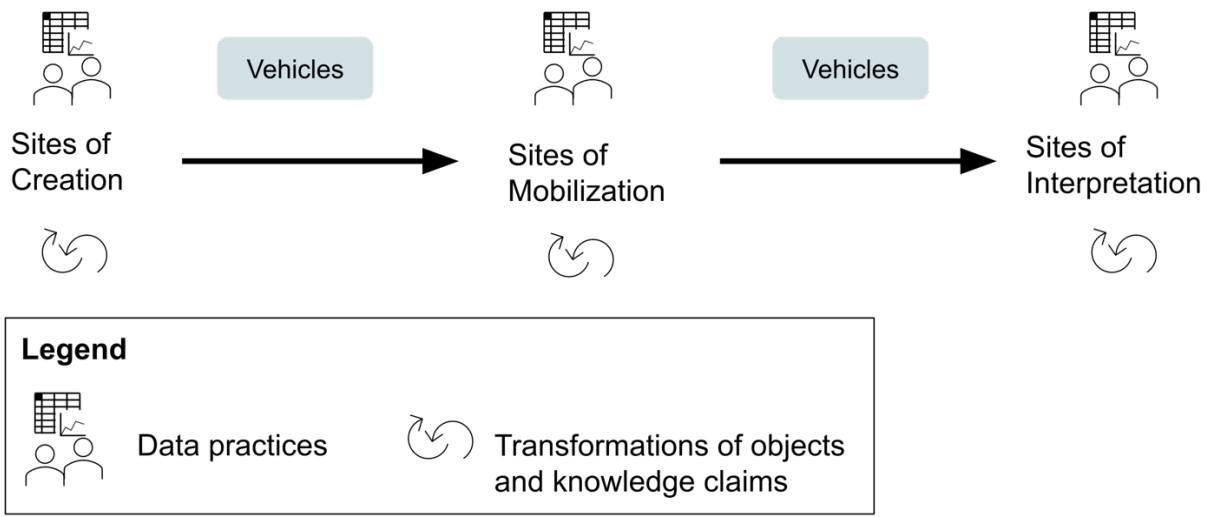

**Figure 1.** Components of the conceptual framework showing the broad dynamics involved in data journeys in the popularization of science.

Although the term 'data journeys' has been used in various literatures (Bates et al., 2016; Daga & Groth, 2023), we adopt the conceptualization outlined by Sabina Leonelli and colleagues (Beaulieu & Leonelli, 2021; Leonelli, 2016; Leonelli & Tempini, 2020). In this view, data journeys are composed of a "myriad of techniques, efforts, instruments, infrastructures and institutions used to mobilize and process data" (Leonelli & Tempini, 2020, p. v); these elements work together to help data to travel across space, time, and social contexts.

Data journeys can be classified into particular stages of *data practices*, including data collection, cleaning, and visualization, which occur across sites of creation, mobilization, and interpretation (Beaulieu & Leonelli, 2021, Figure 1). Data are altered to fit new environments as they travel, e.g., by getting new labels or through merging and deletion (Leonelli, 2020). These modifications allow data to be made usable by new audiences for different purposes, which in turn shapes the meaning and values which data carry. Such modifications also raise questions about when a dataset stops being the same dataset and starts being something else. Focusing on the narrative of a *data journey* rather than seeing data as a single entity provides a more holistic view. Studying data journeys also requires thinking about *how* data travel, by examining the 'vehicles' that transport data on their journeys, e.g., documents and case reports (Ankeny, 2020), and the relationships between data and the various actors who are involved in making data portable, from scientists to data curators to journalists.

The concept of data journeys aligns with research on the popularization of scientific knowledge (Hilgartner, 1990). Critiquing the 'dominant model' of science communication, Hilgartner emphasizes that the spread of scientific knowledge is not a linear transfer of genuine science into appropriately simplified versions fit for public audiences. Instead, scientific knowledge is transformed as it moves across formats and modes. These transformations are often multi-directional, such that scientists, mediators (e.g., journalists), and audiences all contribute to how science is constituted (Felt & Davies, 2020).

Throughout processes of popularization, *transformations* therefore occur in both scholarly products and knowledge claims (Latour & Woolgar, 1986, Figure 1). These transformations happen within academic processes, e.g., when researchers teach or write grant proposals, or in other settings, e.g., when journalists write news stories or create data visualizations (see also Brennen, 2018). Importantly, the dominant model — the idea that it is straightforward to distinguish between 'genuine' science and its 'popularized' versions — privileges the viewpoint of scientists, as researchers have the power to decide what constitutes an appropriate simplification and to denounce communication which does not match their own viewpoints (Hilgartner, 1990).

## 3. Case description and methodology

Founded as a weekly publication for "the ambitious youth of the machine age" in 1845, *Scientific American* (*SciAm*) is the longest continuously-published magazine in the English language (Mott, 1930). Since then, the magazine's envisioned readership, content, and form have evolved. In our interviews, *SciAm* described their current readership as an educated science-interested public and sees the publication as one of few magazines (along with National Geographic and Science News) that provides in-depth science reporting in a way that is understandable to curious lay people. Online readers tend to be younger, more globally distributed, and are seen to have a desire for higher-level coverage of topics. Classic stories in the magazine have covered topics in space and physics, life and environmental sciences, and paleontology. An increasing amount of coverage explores how science intersects with social issues.

A special issue commemorating the 175th anniversary of the magazine demonstrates in more detail how *SciAm* articles have evolved over time (Scientific American, 2020). An analysis of the length and structure of articles found a decrease in the number of characters per article after a magazine redesign in 2001. This shift is attributed to more room being allocated for visual elements, e.g., photography and special typography (Stefaner et al., 2020). Today, the magazine publishes in-depth feature articles, news stories, expert opinions, and other media, both print and online, reaching millions of people each month[1]. At the time of our study, *SciAm* employed

---

[1] https://www.scientificamerican.com/page/about-scientific-american (Retrieved in June 2023)

approximately 30 people, including six news editors, six graphics employees, and six production staff, along with various managers.

To answer our research questions about producing data visualizations and the affordances of open data at *SciAm*, we used a mixed methods approach incorporating interviews and visual analysis.

### 3.1. Interviews

We conducted 11 one-hour semi-structured interviews via videoconferencing with current and past members of the *Scientific American* staff. We used snowball sampling to recruit participants involved in chart creation or in decisions that could influence how data are visually communicated. Sampling of participants stopped once we had spoken with representatives from all professional roles involved in visual data communication who were willing to participate in the study. This represented the overwhelming majority of people involved in producing data visualizations and data-driven stories on our topics of interest. We also determined that we reached sufficiency in the interviews, i.e., that we had collected enough data to address our research questions (LaDonna et al., 2021; Vasileiou et al., 2018). Participants worked as text editors, graphics editors, visualization designers, fact-checkers, and in leadership positions. Seven participants had over 15 years of experience in their roles; three reported having 6 to 15 years of experience; and one participant had been working at *SciAm* for one year. To protect the anonymity of participants, we describe their demographics at an aggregate level only. In our findings, we attribute quotes to participants using a number (**PX**) to differentiate between viewpoints.

Prior to the interview, participants selected at least one article that included at least one visualization that they had worked on about climate change or COVID-19. During each interview, we asked participants to walk us through the process of producing the article and its visualizations. Examples of these visualizations are included in the findings as 'visual vignettes' in Boxes 1-3. We asked further questions about collaborative (data) work practices, design and aesthetic choices, editorial guidelines, and audience engagement. Questions were tailored to match participants' expertise and professional roles. The interview protocol is available in the supplementary material. Interviews were recorded, transcribed, and then coded using ATLAS.ti (ATLAS.ti Scientific Software Development, 2022).

We used a combination of deductive and inductive coding and analyzed the coded transcripts using reflexive thematic analysis (Braun & Clarke, 2006, 2019). The codebook went through three rounds of iterative development with two of the authors. Deductive codes were based on the literature informing our conceptual framework and the structure of our interview protocol, with codes such as audience engagement, story choice, and data access. Inductive codes focus on particular types of data work, transparency, and misinformation. The first author coded all transcripts and consulted with the second author when clarity was needed. This study is part of a project that has undergone ethical screening according to the guidelines of our institution and has been determined to be low-risk.

### 3.2. Visualization analysis

We draw on a visualization analysis first reported in our earlier work (Koesten et al., 2023) to supplement our interview findings, which are our main data source in this paper. In our previous study, we conducted a characterization and thematic analysis of data visualizations published in articles about climate change and COVID-19 (and pandemics more generally) in *Scientific American*, published between 1970 and 2022. This corpus is further described in Table 1.

**Table 1.** Corpus description. A single article can contain multiple data visualizations.

|  |  | **Articles (n=58)** | **Visualizations (n=171)** |
|---|---|---|---|
| **Subject** | Climate change | 49 | 135 |
|  | COVID-19 | 7 | 34 |
|  | Pandemics generally | 2 | 2 |
| **Type** | Print | 35 | 107 |
|  | -   In-depth feature articles | 27 | 97 |
|  | -   One-page graphic-centered articles | 8 | 10 |
|  | Online (Brief news-oriented pieces) | 23 | 64 |
| **Publication year** | Between 2017–2022 | 43 | 58 |
|  | Between 1970–2016 | 15 | 59 |

Here, in a new analysis of the corpus, we focus on codes pertinent to answering our research questions about data visualization production practices and open data and to our conceptual framework. Codes used in this analysis include descriptions of data sources and references, data visualization producers, article author information, and article subject information. During the visualization coding process (described in Koesten et al., 2023), we also created detailed memos documenting our questions, challenges, and observations. We also asked *SciAm* to comment on their goals for creating selected visualizations, including the visualizations presented in Box 1-2, as described in (Koesten et al., 2023). For the analysis reported in this paper, we thematically coded these memos and comments, informed by our conceptual framework and research questions. We further examined the magazine's style guide, provided by *SciAm*. We interweave the results of the visualization analysis into our narrative interview findings to provide a more contextualized analysis.

## 4. Findings

We begin by exploring the data and work practices that constitute *data journeys* at *Scientific American* (Section 4.1) before focusing on the *transformations* that occur in data visualizations during the popularization of scientific knowledge (Section 4.2). Finally, we examine how data, text, and visual elements are crafted into new *vehicles*, enabling data to travel to audiences and sites of consumption in future journeys (Section 4.3, Figure 2).

### 4.1. Tracing data journeys: Data flows and workflows

The decision-making processes involved in making data travel at *Scientific American* occur in different iterative phases of work, described below.

#### *Deciding on which story to tell*

At the beginning of a data journey at *SciAm*, staff members come together to propose and discuss possible stories to further develop for publication. Story pitches (and data) arrive at the magazine via various mechanisms, or 'vehicles' which bring data to *SciAm* (Figure 2). Editors and journalists hear about new academic work by monitoring academic conference circuits, publication alerts, social media accounts, other news outlets, and university press releases. Scientists and journalists also pitch stories directly to *SciAm*, attracted by the reputation of the magazine; editors, in turn, know scientists working on specific topics and reach out to them directly. This can create a Matthew effect (Merton, 1968), where work that is already being written about, publicized, or published by known researchers is further disseminated.

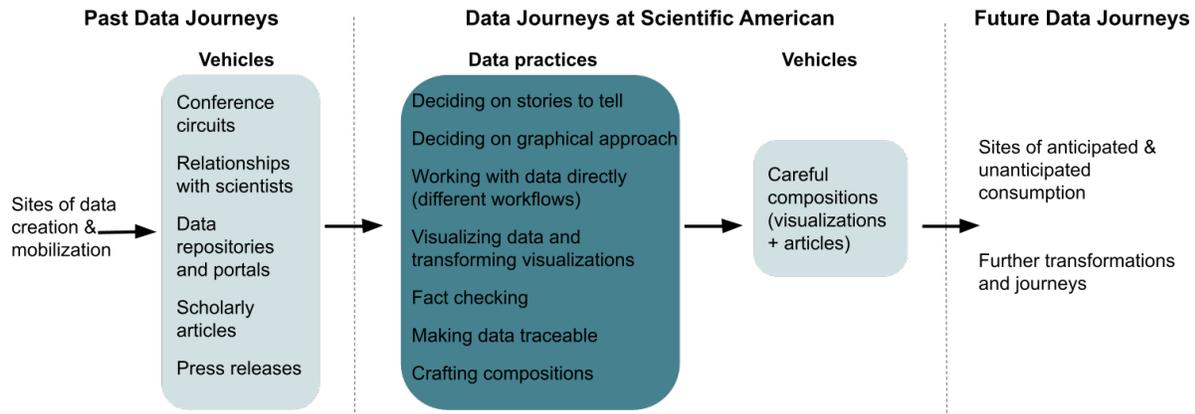

**Figure 2**. Data journeys at SciAm involve practices (dark teal) and vehicles (light teal) used to make data travel to new sites.

*SciAm* also chooses story ideas based on what is interesting to staff members, as educated 'science-minded people,' which is also how they describe their intended readership. Ideas can be born from conversations or events in staff members' personal lives; these ideas are brought to pitch meetings where editors collaboratively decide what will be published, acting as brokers between science and different publics (Allen, 2018). Decisions can be based on tacit knowledge, developed over years, about what will be interesting and emotionally engaging for readers. Editors use their own emotional responses as an initial gauge to estimate the reception of a story.

> *There's the data, but there's an infinite number of stories we could tell [..] If we have an emotional response, that's kind of the first level of oh, maybe this is something that people will also respond to. (P11)*

Staff members also choose stories as a way of drawing attention to concerning topics that have the potential to be collectively solved. Staff do not see themselves as activists in a political sense, but they care deeply about stimulating change and providing context, for both climate change and COVID-19.

> *Wherever we can sort of put it in that context of saying [..] here's something we can do that can make a difference, we try and hit that note. I think it helps people feel less paralyzed. The reason we write about climate change is because we think it's important and something we should be doing something about. We don't think of ourselves as advocates in any political sense, but obviously, as someone who has written about this for so long, I really want us to do something about climate change. (P5)*

Providing context about data and numbers is seen as an important mechanism to counter common misperceptions and misinformation. This desire to fight misinformation and reach new audiences, especially in the case of climate change, means that editors decide to cover the same old story in different ways. A particular story may be new for some readers, but longtime subscribers may have previously encountered the content. Much climate change research also focuses on similar topics, e.g., temperature increases or rising sea levels. We observed this in our article corpus, where topics such as weather extremes, greenhouse gas emissions, or past and future warming predictions recur across years of publication (Figure 3). Editors search for stories that have new insights about recurring topics, aiming to develop stories that emphasize the problem but avoid "beating them [readers] over the head" (P3).

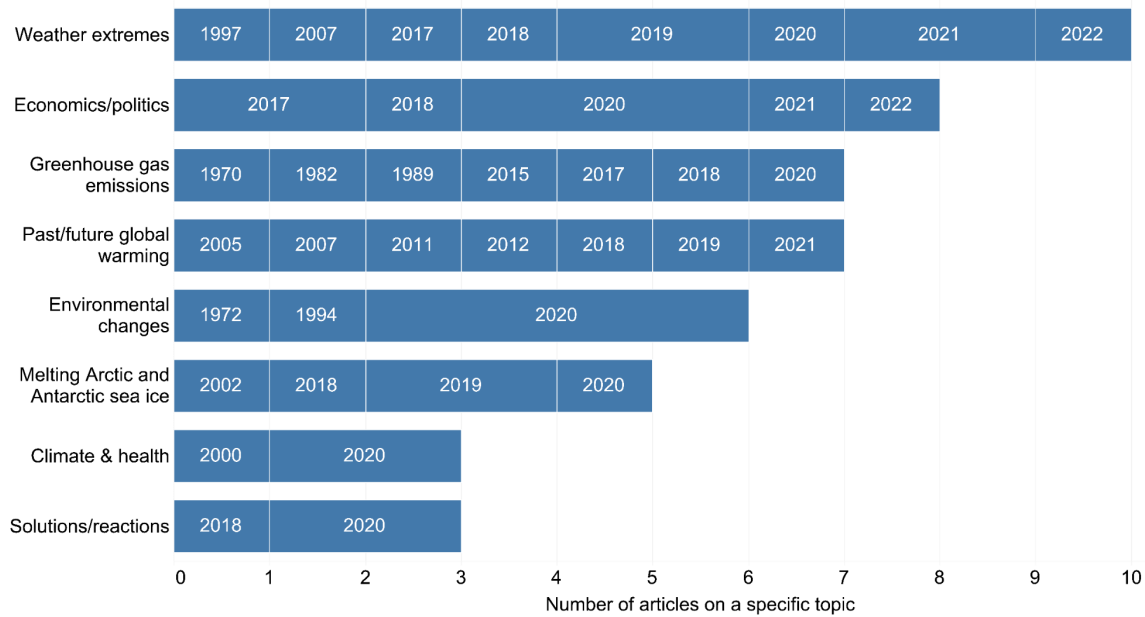

**Figure 3.** Topics present in the climate change articles in the analyzed corpus (n=49). Bar length represents the number of articles per year on a specific topic.

### *Deciding on a graphical approach*

After a decision has been made to pursue a story, *SciAm* staff begin to plan their graphical approach. Staff decide if a story should be graphics-focused, meaning that it will incorporate data and data visualizations, or if photographs or bespoke illustrations and art will accompany the text. Data visualizations are just one part of the visual plan of a story; each visual element has affordances about what it can communicate. According to one editor,

> *If it's about elephants, I want to see the elephants. If it's about the James Webb Space Telescope, we want to see the images from the Space Telescope. But some stories, you know, like about Alzheimer's disease or other things, there aren't good photos (P11).*

Similarly, including data visualizations is not appropriate for every story, nor are all types of visualizations; data can "strengthen a story, or it can weaken it, depending on how it's used" (P8). Graphics staff, editors, and journalists develop dynamic visual plans for how to incorporate visual elements into an article. This involves looking for the graphics potential of a particular story, commissioning art from freelancers, and engaging in back-and-forth conversations with each other and researchers. Each team member brings their own expertise to these exchanges; the development of the visual plan can involve simultaneously working with data and experimenting with different visualization types.

### *Working with (open) data*

There are two primary workflows for how *SciAm* staff work with data; these are informed by the type of reporting being done and the availability of the data. In the first workflow, which we term the 'narrative workflow,' *SciAm* staff know the narrative and message of a data visualization and story in advance. They have usually already encountered the analyzed data (e.g., via a known source, an academic article, or through a scientist), and they need the data to make a point in the story. While they may or may not have already seen a visualized version of the data (created by the scientist) in this workflow, they know the scientific conclusions the data have been used to support.

*Most of the articles I work on come out of the text first, especially from scientist authors, or journalists who propose or pitch a story idea. Then we develop graphics that support that story (P1).*

The visual vignette presented in Box 1 is an example of a visualization produced using this workflow, which is the more common data workflow at the magazine. Although the design of this visualization was significantly different from the original, the message and narrative of the overall visualization remained the same as the conclusions drawn by the scientist, underscoring the power that scientists exert in the popularization process (Hillgartner, 1990).

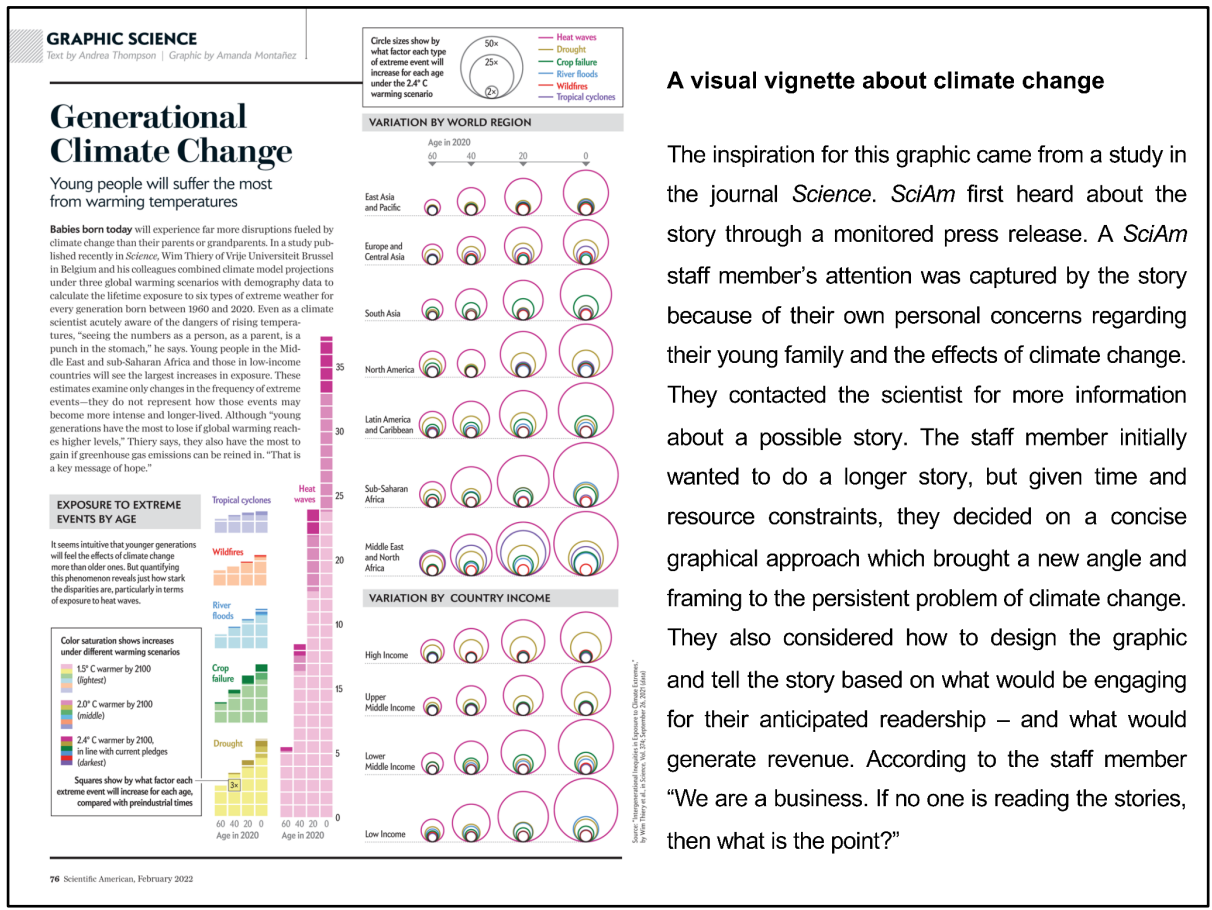

**Box 1**. A visual vignette describing the production of a climate change visualization. Reproduced with permission. Copyright © (2022) *SCIENTIFIC AMERICAN,* a Division of Springer Nature, America, Inc. All rights reserved. Originally published in (Thompson & Montañez, 2022).

In the second workflow, which is similar to the 'data-driven path' proposed by Parasie (2015), *SciAm* staff work using a 'light hypothesis' in an exploratory approach. They have questions of their own and seek data to investigate them, as in the case of the visual vignette presented in Box

2. Journalists track down the needed data to explore, e.g., the effects of the pandemic on the funding of clinical trials. This workflow gives *SciAm* staff and hired freelancers the freedom to discover new insights or to change direction as they uncover or collect new data. It is a constant process of questioning, researching, and looking for data — and pivoting if the data do not exist or are not findable.

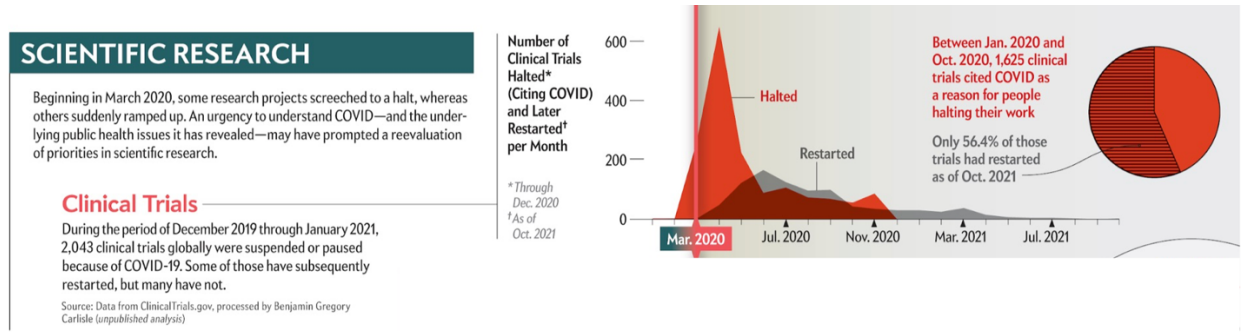

**Box 2.** A visual vignette describing the production of a COVID-19 visualization. Reproduced with permission. Copyright © (2022) *SCIENTIFIC AMERICAN,* a Division of Springer Nature, America, Inc. All rights reserved. Originally published in (Montañez & Christiansen, 2022).

In both workflows, *SciAm* works with data directly to create visualizations. Where an open dataset is available, this is done by downloading it from a repository or data portal; in other cases, *SciAm* requests data from scientists. In the past, before data were shared as openly, *SciAm* would transcribe data from existing charts in the academic literature if necessary, although this is not done currently. This type of transcription is a practice in which scientists also engage (Gregory et al., 2019). When necessary, *SciAm* staff ask scientists questions to help understand data encountered in academic literature. In the second workflow, it is not always possible to consult with a scientist, leaving staff to rely on the provided documentation to understand the meaning of variables, such as 'person year' in government data.

There is also a middle ground between these two workflows, based on *SciAm*'s tacit knowledge about the general conclusions made across scientific studies. Here, they do not highlight a particular piece of research but rather look for data that will help them to tell the story that is emerging and becoming accepted within academia. They are operating from a more hypothesis-driven approach (Parasie, 2015) as they dig into the data to see if broader stories, trends, or hunches they have observed are supported.

> *Sometimes we have anecdotal ideas that something has occurred or that there's been a trend. Then you get the data, and you go, oh, well, that's not actually as clear as we thought it was, or maybe we're really [only] talking about a subset of what the data is showing. And that gives us an opportunity to go into the weeds a little bit and talk it out a bit. (P10)*

### *Creating the visualization*

Working directly with the data to create visualizations allows *SciAm* to own the copyright to graphics they produce; it also allows them to potentially avoid mistakes by reproducing errors present in previous publications or visualizations and gives designers and artists more freedom to craft something unique, creative, and engaging. Visualization designers make decisions about what data to display and where to put the focus of a visualization. This often boils down to determining and formulating messages for each visualization in advance (further detailed in Koesten et al., 2023); defining messages also helps to guide collaborative data work. Crafting data into a visualization requires a host of decisions, e.g., about chart type, the use of color, and the placement of text on charts. These more visible decisions are just the tip of a longer, iterative process involving transformations of visualizations and carefully composing different elements, as described in Sections 4.2 and 4.3.

### *Fact-checking, review, and accuracy*

Internal rounds of fact-checking begin once draft visualizations are available. Fact-checkers review different versions of visualizations, e.g., for web and mobile interfaces and print editions. This involves double-checking calculations and formulas in spreadsheets, comparing datasets and visualization drafts, 'eyeballing' the size of visual elements, and checking consistency with textual elements. Most of the time, fact-checking involves making comparisons between the visualizations and the datasets and sources provided by the graphics specialists. Fact-checkers also compare the provided datasets and the visualized data to other data sources and their own knowledge. Sometimes the process requires extra attention, especially when working with dynamic datasets that are continuously updated, as in the case of COVID-19 data. Data could change between when the visualization and text were created and when they were reviewed. In the past, *SciAm* needed to work backward from visualizations and reconstruct the data. As data-sharing norms have changed, this is not as necessary, particularly when scientists can review visualizations for accuracy.

> *When I first started, we were doing a lot of tracing and reconstructing data…there was a lot of backward architecture going on. But now I'm finding that scientists are much more open with their data. Especially when they know you're going to be running the final visualization past them for an accuracy check to make sure that you're interpreting things correctly… In the case where we're asking somebody directly for their dataset, involving them and making sure that we're interpreting it properly is pretty critical. (P1)*

### *Making data journeys transparent and traceable*

Data journeys are made visible using references or citations to datasets, data sources, and articles. While *SciAm* has a history of citing its sources and data, *how* they do so is in a state of development. Most commonly, *SciAm* staff reports referencing an academic article where the data were first analyzed. The article provides a gateway for others to access the data, independent of the data's location.

> *Either it might be in the article itself or the supplementary information, but this data is available through that publication in some way. They might need to click through a little bit, but somebody who is saying oh, I want to look at this data myself, if they go to that article, they can find it through a few steps. (P1)*

Referencing a peer-reviewed academic article also allows *SciAm* to back up the rigor of the messages and claims in their own visualizations, as these reflect the findings present in the original publications. When *SciAm* works with data from a portal that has not been 'pre-analyzed' in a scientific publication, they cite the data source itself. They believe that this communicates that their presented analysis has not been peer-reviewed.

> *For somebody who's really interested in deconstructing what we've done here, they can kind of start to suss out what's been peer-reviewed and what's raw data that somebody hasn't necessarily analyzed. It's still perfectly valid data. But…there haven't been necessarily as many eyes making sure that it's been analyzed correctly. (P1)*

References and bylines are also used to recognize the work that is involved in creating graphics. One past member of the graphics team stated that it was not possible to credit all of the work involved in making visuals and putting an issue together; in an ideal world, they would like to add a long note detailing who had been responsible for which piece of work, e.g., "layout by PX, type design by PY" (P7). In contrast, our visualization analysis shows that most bylines include the name of a single graphics specialist.

In our visualization analysis, we found that 88% of the articles in our corpus included some type of reference to at least one data source. Supporting the interview findings, the majority of these data sources were academic publications, followed by references to institutions and organizations, reports, and individuals, both with and without academic affiliations (Figure 4). We classify data that have been analyzed in academic publications or university reports or that are associated with individual researchers at particular universities as data that are likely to have passed through research settings; 42% of the 142 mentions of data sources in the corpus match this criterion. Similarly, 40% of sources were associated either with the name of an intergovernmental or governmental organization, data portal, or report.

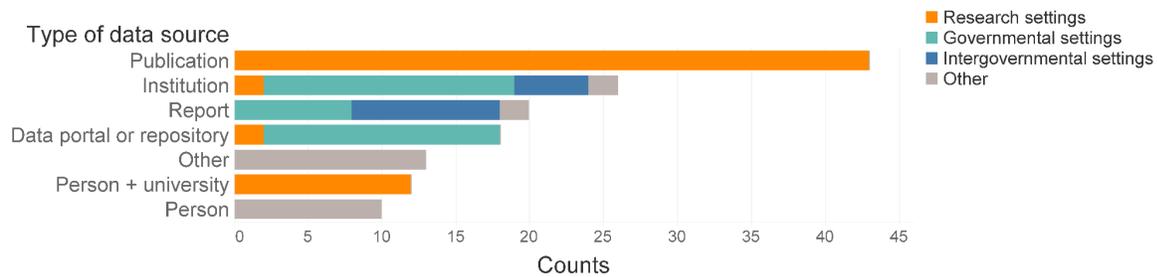

**Figure 4.** Types of data sources mentioned in the corpus (n=142) and the types of settings through which data pass. A single reference could include multiple mentions of data sources.

*SciAm* sometimes integrates data from multiple sources to create a single graphic. Sixteen of the data references included multiple data sources. For example, as described in Box 3, heat wave data from an academic publication were combined with rainfall data from CarbonBrief — a climate website, which itself aggregates data from peer-reviewed studies– and climate data from an intergovernmental report. In almost all of these multi-source references, *SciAm* clearly marks which data in the visualization comes from each source.

As a whole, references to data provide context and facilitate transparency and the possibility of tracing data to their original sources; however, many of the stages in a data journey discussed in this section are not visible in such references alone, including the collaborative nature of producing charts.

## 4.2. Popularization processes: Transformations in visualizations

Thus far, we have described the data journeys through which data enters *SciAm* and become the focus of different work practices. Data visualizations also undergo many transformations through these workflows, which we describe below.

### *Transforming visualizations from existing publications*

*SciAm* staff nearly always make some alterations to visualizations previously published in academic articles. A straightforward time series chart may look quite similar in *SciAm*, but there are always slight changes, e.g., stripping out jargon and adding narrative framings. The process of transforming an academic visualization to a narrative visualization involves communicating with scientists to ensure that *SciAm* maintains the correct message and has interpreted the data appropriately (see Section 4.1). This review is particularly important when *SciAm* staff work with unfamiliar data and topics. While designers value scientists' opinions about accuracy, they are not that interested in hearing opinions about visual design and aesthetics. Scientists sometimes love the transformed versions of their original visualizations; at other times, they are frustrated, especially when (academic) design conventions are broken.

Just as when choosing stories, *SciAm* staff try to transform commonly circulating visualizations to show something new. This is apparent when transforming graphics in reports from the IPCC, as the timely nature of the information puts pressure to create something new (and newsworthy) in a short timeframe. As seen in Box 3, having data readily available — and being able to work with that data — allows *SciAm* to add to the IPCC story in a new way.

**A visual vignette about transforming IPCC charts**

This visualization was created in response to the release of a report from the IPCC. *SciAm* accessed data from an academic publication as well as the IPCC report to create a new spin on the graphics in the widely circulated report. As summarized by a staff member: "The IPCC did their big report on 1.5 degrees C, and we actually did a graphic and a story comparing a few different [scenarios] like sea level rise and a few other things between 1.5 degrees C and two degrees C… This wasn't just 'here's what the report says'… it was simple enough as a graphic that it didn't take as much time versus the graphic science with the younger generations [in Box 1], where that's a lot of data to work with and we need to figure out how to show it. Just showing your sea level rise with 1.5. C versus 2 C is much much simpler" (P5). The simplicity of the design and the story enabled *SciAm* to produce a graphic within a tight timeline, to match the release of the IPCC report.

Credit: Amanda Montaez; Sources: "Extreme Heat Waves under 1.5 C and 2 C Global Warming", by Alessandro Dosio et al., in Environmental Research Letters, Vol. 13., No. 5; April 25, 2018 (heat wave data); Carbon Brief, "The Impacts of Climate Change at 1.5 C, 2 C and Beyond"; accessed October 12, 2018, https://interactive.carbonbrief.org/impacts-climate-change-one-point-five-degrees-two-degrees/# (rainfall data); Intergovernmental Panel on Climate Change, "Special Report on Global Warming of 1.5 C"; October 8, 2018 (arctic, habitat loss, and coral reef data)

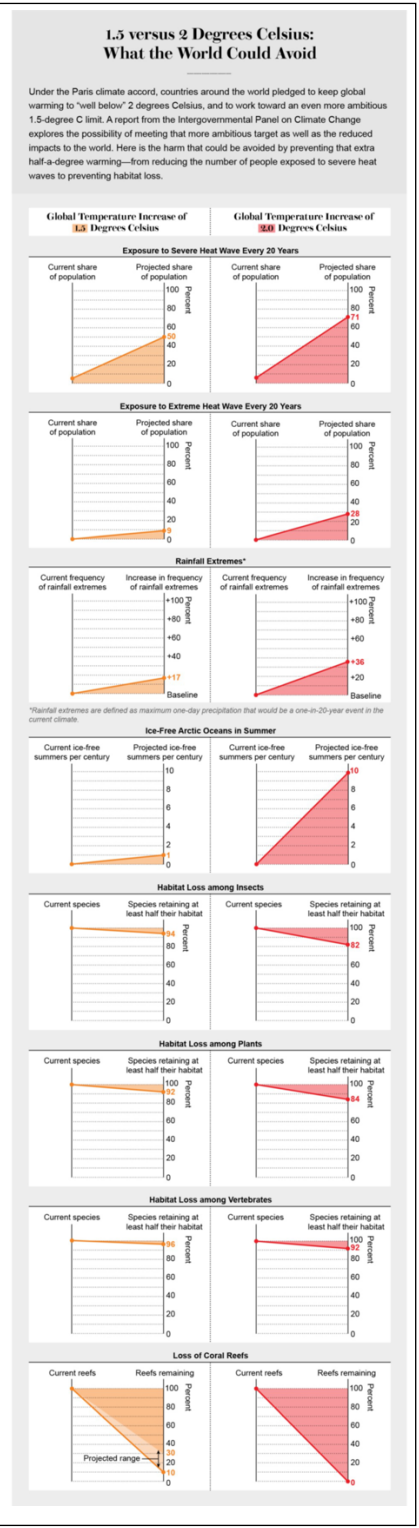

**Box 3.** A visual vignette describing the production of a report based on the IPCC report. Reproduced with permission. Copyright © (2018) *SCIENTIFIC AMERICAN,* a Division of Springer Nature, America, Inc. All rights reserved. Originally published in (Thompson, 2018).

### *Print and online transformations*

Visualizations are also transformed from print graphics into online versions. *SciAm* staff rescale and invert print charts, at times putting the x-axis at the top rather than the bottom to better match online scrolling behavior or to make charts more readable on mobile devices. It is not always possible, or desirable, to transform print visualizations into online graphics. Some transformations would require too much time or effort to provide the same experience online as in print. Instead, *SciAm* works to maximize the impact of print graphics by leveraging the structure of the magazine, e.g., by adding a timeline running through a 6-page spread on the impact of COVID-19 to provide different entry points to the data (see Box 2). The online environment does not automatically equate to a deeper engagement experience:

> *It's more engaging to look at it in print, you can kind of scan along and pick out what you like and don't like, or what you're curious about [..] and this [online version], you're really just scrolling through, and so you might miss something, and or get a little bored [..] It's definitely a challenge to translate that [the print version] when we don't have the right online capabilities for publishing it in a similar way. (P10)*

Online visualizations are often simpler than their print counterparts. Designers use more basic chart types and boil data down into more succinct transformations to match format constraints and the online magazine's envisioned readership. *SciAm* focuses on creating stories and visualizations that provide online readers with "news you can use" (P8). Graphics in print feature articles tend to be more experimental or "mind-bending" (P1), offering a reward to print readers who are believed to enjoy a more challenging reading experience than online readers, who are seen as preferring content that is crisp, clear, and fast to grasp.

### *More than just simplification*

As visualizations are transformed, they are not just simplified. Transformations are crafted to add aesthetic and textual elements, which trigger emotions in ways not encouraged in scholarly communication. *SciAm* carefully considers how to responsibly use color and form to evoke emotions or moods, such as using shades of blue to trigger feelings of somberness in a visualization about death or reds in visualizations about increasing temperatures. They recognize that *SciAm* is not required reading for most audiences and that a more dramatic aesthetic can draw in more readers and improve the magazine's bottom line (Box 1). Aesthetics are also seen as a way to invite people to engage more deeply with data and to reinforce scientific messages in visualizations. Experiments with form and text are also used to reframe climate change stories and data to connect with readers with different needs:

> *It's like, let's try it and throw up the graphic novel form to engage somebody where the focus isn't on the chart. The chart is there to help support this narrative. So I think that's an area where we're trying to meet different people's needs in different ways, in part because there's such a sense of urgency. It's like, 20 years of these targets haven't been making enough of an impact. How else can we present this information in a way that connects with people? (P1)*

### 4.3. Creating new vehicles: Careful compositions for (un)anticipated consumption

*SciAm* crafts new data 'vehicles,' composed of data, visual, and textual elements, preparing data to continue their journey to *SciAm's* readership (Figure 1). Visualizations are rarely "just graphs" (P10) but integrate textual and visual elements to tell a particular story. While it is possible to understand a visualization or text on its own, these elements work best when presented together with other art, images, and text in the magazine. Doing so helps keep readers' attention but also provides moments of relaxation — a much-needed break as readers work through meaty content. The data vehicles created at *SciAm,* therefore, consist not just of data visualizations but also of articles and the magazine as a whole.

*SciAm* considers the affordances of multiple individual elements and how they work in relation to each other in their designs. Graphics can be used to grab readers' attention and advance a story in ways not possible through text; the positioning of data visualizations in an article and the function they play in communicating the larger narrative message depend on what *SciAm* believes will be most useful to readers.

> *Sometimes we're using graphics [..] as connective tissue to help pull people through. Sometimes we're using them as ways of providing background information in a way that allows the text narrative to continue its flow but provide background information that could be useful to some people. And other times, we're using the graphic to tell the story itself. (P1)*

As also seen in their decisions about which stories to tell (Section 4.1), the magazine's envisioned audience — including their imagined interests and ways of consuming content — informs staff's choices. These conceptions of the audience, and by extension, how *SciAm* transforms and visualizes data, are informed by online metrics (e.g., clicking and scrolling behavior), emails from readers, and staff members' own intuition and tacit knowledge about their readership which they have developed over time.

New data vehicles are designed to withstand potentially problematic future data journeys. *SciAm* staff considers how visualizations may travel if others repost them on social media. They embed explanatory text into charts, as observed in our visualization analysis. Annotations, labels, explanatory boxes, and other textual elements were incorporated into 54% of the visualizations in our corpus. This awareness that visualizations may be shared out of context is also reflected in *SciAm*'s internal style guide, which contains instructions for placing "on-art" annotations and labels to explain the significance of different graphical elements (e.g., relationships between coordinates in scatter plots or the use of shading to indicate uncertainty ranges in line charts). Different guidelines exist for different visualization types (e.g., bar charts, small multiples, lollipop, or line charts) but also for the various environments in which they will be displayed (e.g., print vs online, desktop vs mobile).

Carefully portioning data into separate elements is another strategy for providing context and countering potential misuse. Splitting data into multiple visualizations constrains the messages that can be communicated if a visualization is taken out of context.

> *We originally had another graphic that showed the total number of deaths among the vaccinated versus the unvaccinated [..], but we were worried that would be taken out of context, and someone would tweet it and say look at all these vaccinated deaths - which is exactly the opposite of the message we were trying to convey. So we ultimately decided to [..] just have the graphic show the total number — the total vaccinated versus unvaccinated population in this first graphic. And then the second graphic [..] was the death rate. (P8)*

Portioning data in this way steps readers through often complex information and guides them through the composition of an article. In such cases, *SciAm* creates a natural flow where data are mentioned in text before being shown in a visualization, which readers can then examine in more depth. This allows the magazine to more tightly control the narrative and some aspects of future data journeys. It also gives *SciAm*'s envisioned audience succinct arguments they can use to counter misinformation circulating in their own local spheres.

> *Sometimes, we're trying to create stories that you could imagine sending to a friend [or relative] to say, hey, look [..] Because many people might themselves accept vaccines, but they have relatives who don't. And this might be one way to [..] arm them with some data to say, okay, look, vaccinated people can get COVID and die from it, but look at the rates. (P8)*

Data visualizations, particularly those published online, are framed in light of potential science skepticism and mistrust. COVID-19 stories, e.g., are constructed to emphasize the evolving nature of pandemic data — to communicate that "this is what we know *now*, rather than this is what we know" (P8) — to avoid skeptics calling the content incorrect should data change. *SciAm* staff also make careful choices about how to frame climate change articles and visualizations, which are fraught with political tensions that may affect how readers interpret them:

> *Obviously, there's a lot more politicization around climate change. It's a very fraught topic. [..] People have very strong and very entrenched views. [..] It's the milieu that the stories are coming out in, and that can affect reader response. You know, it's the context that they are being read in (P5).*

## 5. Discussion

Taking *Scientific American* as a case study, we analyzed how data visualizations are produced in a popular science magazine. We rooted our analysis in the concept of data journeys and notions about the popularization of scientific information. This allowed us to identify and discuss the practices and visualization transformations involved in bringing data from sites of creation to broader readerships. We conclude by discussing how (open) data and visualization production practices allow, shape, and inhibit data journeys in the popularization of science along four lines:

i) the affordances of open data; ii) the role of collaborative work; iii) misinformation; and iv) transparency.

## 5.1. Affordances of working with (open) data

Our results show that *SciAm* has an established practice of working with data directly to create visualizations. They do this in part to maintain ownership and copyright of the graphics, as does the journal *Nature* (Krause, 2017), another publication owned by *SciAm*'s parent company. As documented in Section 4.1, working with data involves a series of time- and resource-intensive workflows, collaborative work, and skills. Arguably, not every popular science magazine has the same resources that we observed at *Scientific American*. This raises questions about the sustainability of these data practices, particularly as the media landscape sees less investment in (bespoke) print popular science publications (Farhi, 2023). As the amount of available data and research increases (Benjelloun et al., 2020; Thelwall & Sud, 2022), questions arise about how a relatively small team can stay up-to-date and make meaningful decisions about which stories to cover.

As shown in both the visualization analysis (Figure 4) and the description of data workflows, *SciAm* frequently works with data that have passed through research settings. This contrasts with an earlier analysis documenting minimal use of research data in visualizations in news sources (Stalph & Heravi, 2021) and underscores how popular science magazines act as unique hybrid spaces between academia and the mass media (Born, 2020). The difference between our results and those of Stalph and Heravi (2021) also highlights difficulties in classifying data according to their places of creation, e.g., as 'research data' or 'governmental data.' Research is also conducted within governmental and industry settings, and academic researchers also make use of governmental data in their work (Gregory et al., 2020). The concept of data journeys helps to identify sites through which data pass rather than relying on difficult and reductive categorizations based on data origin.

Working with data directly affords more than just copyright ownership of graphics; doing so also enables *SciAm* staff to be creative and diverse in their visual designs in ways that are not (currently) possible within the strictures of most academic publications. Our findings also suggest that having access to data may lead to greater diversity in the messages communicated via data visualizations. *SciAm* does not always rely on data that have been 'pre-analyzed' by scientists; rather, they at times seek data to answer their own questions and add new elements to recurring stories (as in Box 2 and Box 3). This use of data to answer new questions mirrors the potential many see for secondary data reuse in academia (Pasquetto et al., 2017) and has similarities with both the hypothesis-driven and data-driven paths of investigative data journalism observed by Parasie (2015).

**5.2. Collaborative work and contingent encounters**

Various actors shape the narratives transported in data visualizations and popular science articles. The stories created with pre-analyzed data at *SciAm* closely mirror those of scientists themselves, signaling the power scientists hold in defining 'appropriate' popularizations (Hilgartner, 1990). To some extent, scientists also implicitly shape *SciAm*'s stories when staff work without pre-analyzed data, as *SciAm* relies on data documentation and metadata, which are themselves shaped by scientific and curation processes (Edwards et al., 2011).

As they work to understand and use data, *SciAm* staff engage in similar strategies as researchers and other data professionals (Koesten et al., 2021), such as contacting data creators directly for clarification and collaborative problem-solving with colleagues (Koesten et al., 2017; Yoon, 2017). Collaborating with scientists is a key strategy for *SciAm* to ensure the accuracy of their own analyses and visual representations of data. As Pasquetto and colleagues suggest, data creators tend to have an 'intimate and tacit knowledge' about their data that can be traded within collaborations for mutual benefit (Pasquetto et al., 2019). In this case, collaborations between *SciAm* and scientists improve the accuracy of data visualizations, which is in both parties' best interest. Scientists may also benefit from these collaborations by gaining a wider audience for their work or by being able to demonstrate its 'societal impact,' as is increasingly required in the assessment of academic careers (Bornmann, 2013).

While input from scientists is valued at *SciAm*, the magazine makes its own choices about aesthetics, visualization design, and how different elements are pieced together to engage (and increase) its readership. Crafting visualizations and articles involves collaborative work among staff members with different expertise (Koesten et al., 2019); these collaborations are often invisible when looking at charts alone. While graphics staff mostly work with data, our results also document collaborative work between visualization designers, writers, editors, and fact-checkers. Data and visualizations are transformed through these contingent encounters, which in turn shape the composition of an article as a whole.

### 5.3. Misinformation

Our analysis shows that data travel in carefully composed 'vehicles' containing specific arrangements of data, text, and visual elements; this is partially done to provide multiple perspectives and increase engagement and understanding (as recommended in Unwin, 2020). In line with the broader literature describing narrative visualizations (Hullman & Diakopoulos, 2011; Segel & Heer, 2010), *SciAm* constructs visualizations and stories with an eye to how it believes readers would like to (or should) navigate content. They also consider possible 'misuses' of visualizations and data, e.g., to make claims opposite from those which are intended (Correll & Heer, 2017; Lee et al., 2021; Lo et al., 2022). While this type of misinformation is not restricted to digital content, information and data circulated on social media are especially prone to being repurposed and taken out of context (Allcott et al., 2019). *SciAm* attempts to counter visual misinformation by using what Zheng and Ma (2022) term 'defensive design methods,' e.g., by embedding and integrating textual annotations into visualizations or splitting data into separate charts.

Focusing our analysis on data journeys makes many interaction points visible where data, visualizations, and stories are transformed and mixed with other types of content. Misinformation can be introduced at these points, intentionally and inadvertently (Lewandowsky et al., 2012). Our findings demonstrate that *SciAm* is well aware of this risk and that they engage in extensive internal and external fact-checking as a countermeasure. In the transformation processes, different types of (visual) information are also introduced, as aesthetic and artistic elements are added to 'humanize' data and make narratives emotionally appealing (Kennedy & Hill, 2018; Peck et al., 2019) or as *SciAm* staff strive to reframe familiar climate change stories in new ways. Here, their aim is not so much to fight the spread of misinformation directly but to use visualizations as a means of encouraging or revealing opportunities for action among readers (Metag, 2020; Schuster et al., 2023), who can in turn fight misinformation in their own circles.

### 5.4. Transparency and future journeys

Packaging data into vehicles implies that data will travel further, although naturally not all data travel to, or from, *Scientific American*. The stories and the data that *SciAm* chooses are usually already 'flowing' within the academic literature, into data portals and repositories, or via posts on social media. Stories are also chosen based on what *SciAm* believes will be engaging, interesting, and sellable to their readership. This suggests that *SciAm*'s anticipated audience also shapes story and design choices and that the flow of data can be understood as co-produced between many actors, including researchers, magazine staff, and readers (Felt & Fochler, 2011).

Once packaged into visualizations and articles, data travel into the world of *SciAm* subscribers and readers but also to other (unanticipated) publics. *SciAm* works to make previous steps in data journeys transparent, particularly by crediting the work of chart producers and citing data sources. As with other data journalists (Kennedy et al., 2020; Zamith, 2019), findings from both the interviews and the visualization analysis indicate that *SciAm*'s data citation practices are in a state of development. We also found that *SciAm* often provides extra context to their data references to increase transparency (e.g., by clearly indicating which data corresponds to which part of a figure) or to indicate if data had undergone prior analysis, as in Box 2. These 'contextual additions' go beyond what is normally recommended in academic data citation standards (e.g., American Psychological Association, 2022; DataCite, n.d.).

*SciAm*'s practices for referencing data sources are problematized by issues in the broader landscape of scholarly and science communication. Data themselves are not always persistently available (Federer et al., 2018; Pepe et al., 2014) or able to be cited at a level of granularity that enables them to be found (Silvello, 2018). Many academic publications, which are commonly cited data sources at *SciAm*, are also not openly available (Piwowar et al., 2018); this hampers the ability of publications to act as a gateway to the data. Despite these wider problems, our results demonstrate that *SciAm* is experimenting with and developing their own practices for referencing data rather than waiting for a 'perfect' solution to be developed, which may not be possible (Lowenberg, 2022; Lowenberg et al., 2021).

As *SciAm* visualizations travel to new contexts, *SciAm* becomes responsible for the data in a new way. They own the rights to the visualizations, although the data come from somewhere else.

This raises its own set of questions about transparency and documentation and emphasizes the importance of having robust, persistent links not only between scholarly publications and data (Silvello, 2018) but also with other forms of science communication.

## 6. Conclusion

Drawing on the concept of data journeys, we explored how data travel to visualizations in popular science magazines, taking *Scientific American* as a case study. We shed light on the mundane and collaborative practices involved in producing data visualizations and discussed the effects of these practices. We also provided a detailed example of how the concept of data journeys can be used as an analytical framework.

Making use of the data journeys concept allowed us insight into how different meanings and connotations become a part of data as they travel. *SciAm* works to maintain the original scientific meaning and contexts present in the data which they work with; at the same time, the connotations of data are changed in both larger and more incremental ways as they journey through the magazine. This is particularly visible in the data-driven workflow we describe, where SciAm works with data not previously discussed in academic articles to answer their own research questions. It is also visible in the aesthetic and design choices which *SciAm* makes as they visualize data with the aim of engaging readers, triggering emotions, or fighting misinformation. These perhaps more incremental changes — be they the stripping of jargon, recoloring of lines, or placement of charts — add layers of interpretation, connotation, and meaning to data and affect how future readers may understand data and use them in their own lives.

Our approach highlights the importance of focusing on the *practices* involved in data journeys and data transformations. From a practical standpoint, paying attention to these practices can be a starting point for identifying where tools and guiding principles to preserve and develop meaning for 'popularized data' could be developed. Examining *SciAm's* data practices also reveals one set of strategies that may be of interest to broader scholarly and science communication communities, particularly regarding visualization design and transparency.

While analyzing a data journey arguably provides a more holistic view of data reuse, it also raises theoretical questions about what makes a data journey a successful journey (see also Leonelli &

Tempini, 2020). Is a journey successful when data get where they are intended to travel? Or when data tell an intended story or are used in an anticipated manner? Perhaps a journey is successful if data travel to many different places or if something is learned along the way? Answering these questions requires further investigation of data practices at future sites in a data journey, from multiple perspectives.

## Disclosure Statement

This work has been funded by the Vienna Science and Technology Fund (WWTF) [10.47379/ICT20065]. The authors have no conflicts of interest to declare.

## Acknowledgements

We would like to especially thank *Scientific American* for their collaboration and participation in this study and Alice Fleerackers for the close reading and insightful comments. We would also like to give special thanks to the Technosciences, Materiality, and Digital Cultures research collective at the University of Vienna for their early input into this article and to Christian Knoll and Michaela Hubweber for their comments and visualization advice.

# Supplementary Material

**Semi-structured interview protocol**

The interview protocol consisted of four main parts, including the collection of demographic information. Prior to the interview, participants selected relevant article(s) with visualization(s) which served as a discussion prompt in Part 2. The protocol was tailored to individual's areas of expertise.

**Introduction**

Thank you for participating in this study and for your time today.
Thank you for completing the informed consent form. Do you have any questions about the form or the interview?

Begin recording of the interview.

Outline goals of the project and the study.

- As you read in the informed consent, this study is part of the Talking Charts project, which investigates how climate change and COVID data are visualized in scholarly and popular science media.

- As part of this project, we are analyzing visual data communication in *Scientific American* on the topics of climate change and the COVID-19 pandemic. This analysis will serve as a case study for how visualizations are used to communicate data about these topics to non-academic audiences.
    - This involves two parts: an analysis of charts and semi-structured interviews.

- The goal of the interview study is to learn more about the production and reception of charts in Scientific American. We will ask questions, e.g., about editorial and visualization production practices, guidelines and the audiences and readership of the magazine.

Overview of the interview
- The majority of the interview will be a conversation centered around one of the articles which you sent in advance.

## Part 1: Demographics

- What is your professional role?
- How long have you been working in this capacity / in the profession?
- What is your (academic/professional) background?
- Could you describe the *Scientific American* publication a bit for context?
    - How do you see the role of the publication in the wider science communication landscape?

## Part 2: Visualization production and data work

*In this section, participants discussed the sample article(s) and visualization(s) which they had sent prior to the interview. The interviewer or participant shared their screen with the visualization/article which was being discussed.*

- Could you walk us through the creation of this article?
- Prompts:
    - Why was this story chosen?
        - Why and how do you generally choose the stories which you do?

- - - Are stories always related to an academic publication?
    - Do you have access to those publications?

  - Why was this story made online vs print?
    - Which stories are chosen for print? And which for online? How are the two related?

  - How do charts and other elements (text, photos, etc.) work together to tell the narrative of the story?
    - Do you see charts as something which can stand alone? Or are they best understood in relation to an entire article?

  - How were the charts made? Who made them? For which purposes?
    - Are charts usually derived from another graphic or a copy of an original graphic?

  - Where did the data come from?
    - Are data available outside of a publication (e.g., in a repository or governmental portal?)
    - Do you access the data directly?

  - Could you describe the experience of using secondary data (data which you don't create yourself)?

  - Could you discuss the role of collaborating with others when creating charts?
    - How do you work with data creators, with scientific experts, with other team members, etc.?
    - Are there any differences here between creating print or online charts?

  - What is the role of aesthetics in chart creation?

  - Would you say these processes which you describe are typical?

**Part 3: Audience-related questions**
- Who is the intended audience of Scientific American? Do you know if the intended audience matches the actual audience?

- How does an envisioned reader work through the article? Do they start with visualizations first or text?

- Do you think that the readership understands the articles in Scientific American? Do they understand the visualizations?

- How do you (or have you) measured audience engagement? (i.e., logs, downloads, web analytics, focus groups) How have these insights been used?

- How do people end up at Scientific American (both print and online)?

**Part 4: Publication-related questions**
- Do you have documented editorial guidelines/procedures?
    - Could you describe those?
    - How are they used? How are people trained?

- What are some of the biggest changes have you observed in Scientific American over time?
    - Are there things which you used to do, but don't any longer? For example, some material which was only online, and now is in print, or vice versa?

**Closing**
- Are any of these processes, etc. which we have discussed today unique for climate change stories or charts? For COVID-19 stories or charts?

- Thank you very much for your time and insight.